\documentclass[useAMS,usenatbib]{mn2e}
\usepackage{txfonts}
\usepackage{graphicx}
\usepackage{multirow}
\usepackage{ulem}

\newcommand{\apj}{ApJ}
\newcommand{\apjl}{ApJ Let.}

\newcommand{\aap}{A\&A}

\newcommand{\mnras}{MNRAS}

\newcommand{\kms}{km~s$^{-1}$}

\newcommand{\heii}{\textrm{He\,{\sc ii}}}

\newcommand{\hei}{\textrm{He\,{\sc i}}}

\newcommand{\msun}{{\rm\ M}_\odot}

\newcommand{\degree}{$^{\circ}$}

\bibpunct{(}{)}{,}{a}{}{,} 

\title[Modelling KT Eri]{Morpho-Kinematical Modelling of Nova Eridani 2009 (KT Eri)}
\author[V. A. R. M. Ribeiro et al.]
	{V. A. R. M. Ribeiro,$^{1}$\thanks{E-mail: vribeiro@ast.uct.ac.za}\thanks{South African Square Kilometre Array Fellow.}
        	M. F. Bode,$^{2}$
	M. J. Darnley,$^{2}$
	R. M. Barnsley,$^{2}$
        \newauthor
        U. Munari,$^{3,4}$ and
        D. J. Harman$^{2}$ \\
        $^{1}$ Astrophysics, Cosmology and Gravity Centre, Department of Astronomy, University of Cape Town, Private Bag X3, Rondebosch 7701, South Africa \\
        $^{2}$ Astrophysics Research Institute, Liverpool John Moores University, Twelve Quays House, Egerton Wharf, Birkenhead, CH41 1LD \\
        $^{3}$ INAF Astronomical Observatory of Padova, via dell'Osservatorio, 36012 Asiago (VI), Italy \\
        $^{4}$ ANS Collaboration, c/o Astronomical Observatory, 36012 Asiago (VI), Italy
        }
\begin{document}

\date{Accepted 2013 May 15.  Received 2013 May 13; in original form 2012 August 30}

\pagerange{\pageref{firstpage}--\pageref{lastpage}} \pubyear{2002}

\maketitle

\label{firstpage}

\begin{abstract}
Modelling the morphology of a nova outburst provides valuable information on the shaping mechanism in operation at early stages following the outburst. We performed morpho-kinematical studies, using {\sc shape}, of the evolution of the H$\alpha$ line profile following the outburst of the nova KT Eridani. We applied a series of geometries in order to determine the morphology of the system.  The best fit morphology was that of a dumbbell structure with a ratio between the major to minor axis of 4:1, with an inclination angle of 58$^{+6}_{-7}$ degrees and a maximum expansion velocity of 2800$\pm$200 \kms. Although, we found that it is possible to define the overall structure of the system, the radial density profile of the ejecta is much more difficult to disentangle. Furthermore, morphology implied here may also be consistent with the presence of an evolved secondary as suggested by various authors.
\end{abstract}


\section{Introduction}
A classical nova outburst occurs on the surface of a white dwarf (WD, the primary) following extensive accretion from a secondary star \citep[typically a late type main-sequence star that fills its Roche Lobe;][]{CK56}. This outburst, a thermonuclear runaway, ejects $\sim10^{-6} - 10^{-4}$~$\msun$ of matter from the WD surface with velocities from hundreds to thousands of \kms\ \citep[see, e.g][and references therein]{BE08,B10}. Nova outbursts can be divided into Classical Novae (CNe) and Recurrent Novae (RNe). As the latter name suggests, this outburst usually occurs on time-scales of order tens of years and the majority are observed to have evolved secondaries \citep[e.g.][]{DRB12}. Such a short recurrence time-scale is usually attributed to a high mass WD, probably close to the Chandrasekhar limit, together with a high accretion rate \citep{SST85,YPS05}.

Nova Eridani 2009 (hereafter, KT Eri) was discovered on 2009 November 25.54 UT in outburst by \citet{I09}. Low resolution optical spectra obtained on 2009 November 26.56 showed broad Balmer series, \hei\ 5016\AA\ and N {\sc iii} 4640\AA\ emission lines with FWHM of H$\alpha$ emission around 3400 \kms\ \citep{MF09}. This object was confirmed later as a ``He/N'' nova \citep{RPR09}.

The discovery date was not the date of outburst. Other authors suggested the nova was clearly seen in outburst on 2009 November 18 and that the outburst occurred after 2009 November 10.41 \citep{DDG09}. \citet{HBH10} searched the Solar Mass Ejection Imager (SMEI) archive and the Liverpool Telescope SkyCamT observations and found KT Eri clearly in outburst on 2009 November 13.12 UT with a pre-maximum halt occurring on 2009 November 13.83. They found the outburst peak to have occurred on 2009 November 14.67 $\pm$ 0.04 UT (which is taken as $t$ = 0) with magnitude $m_{\textrm{SMEI}}$ = 5.42 $\pm$ 0.02. Furthermore, \citet{HBH10} show KT Eri to be a very fast nova with $t_2$ = 6.6 days (the time taken for the magnitude to decline by two magnitudes from optical peak). A distance to the system has been derived by \citet{RBS09} as 6.5 kpc. However, it is noteworthy that this was assuming $t_2$ = 8 days which is slightly different from that derived by \citet{HBH10} (although the SMEI magnitudes are white light effectively and the $t_2$ would be expected to differ from that in V for example).

KT Eri has also been detected at radio wavelengths \citep{OMS09} and as an X-ray source \citep{BOP10}. The {\it Swift} satellite's first detection of KT Eri with the X-ray Telescope was on day 39.8 after outburst as a hard source \citep{BOP10}. This was also the case on day 47.5. However, by day 55.4 the supersoft source (SSS) phase emerged (the dates have been corrected to the $t$ = 0 given above). On day 65.6 the SSS softened dramatically. \citet{BOP10} also noted that the time scale for emergence of the SSS was very similar to that in the RN LMC 2009a \citep{BOP09LMC}.

An archival search for a previous outburst was carried out with the Harvard College Observatory plates by \citet{JRD12}, where they identified the progenitor system. However, no previous outburst was observed. They found a periodicity of 737 days which was attributed to reflection effects or eclipses in the central binary. A shorter periodicity of 56.7 days was also found by \citet{HCW11} which was attributed to arise from the hot spot on the WD accretion disk although \citet{JRD12} find no evidence to support the shorter frequency. \citet{JRD12} also derive a quiescent magnitude $<B>$ = 14.7$\pm$0.4 which combined with the periodicity above suggests that the secondary star is evolved and likely in, or ascending, the Red Giant Branch. This result is corroborated by \citet{DRB12} and \citet{IT12}. Furthermore, \citet{JRD12} predict that if KT Eri is a RN then it should have a recurrence timescale of centuries.

In this paper we present mopho-kinematical modelling, using {\sc shape}\footnote{Available from http://bufadora.astrosen.unam.mx/shape/index.html} \citep{SKW11}, of the evolution of the H$\alpha$ line profile. {\bf } In section~\ref{obs} we present our observational data; in section~\ref{model} we present the modelling procedures and the results are presented in section~\ref{results}. Finally, in section~\ref{discussion} we discuss the results and present our conclusions.

\section{Observations}\label{obs}
Optical spectra of KT Eri were obtained (Table~\ref{tb:tb1}) with the Fibre-fed Robotic Dual-beam Optical Spectrograph \citep[FRODOSpec,][]{MCS04}, a multi-purpose integral-field input spectrograph on the robotic 2m Liverpool Telescope \citep{SSR04}. The dual beam provides wavelength coverage from 3900$-$5700\AA\ (blue arm) and 5800$-$9400\AA\ (red arm) for the lower resolution (R = 2600 and 2200, respectively) and 3900$-$5100\AA\ (blue arm) and 5900$-$8000\AA\ (red arm) for the higher resolution (R = 5500 and 5300, respectively).
\begin{table}
  \centering
  \caption{Log of optical spectral observations of KT Eri with the FRODOSpec instrument on the Liverpool Telescope.}
  \begin{tabular}{lcc}
    \hline\hline
    \multirow{2}{*}{Date} & Days after & Exposure \\
    & outburst & time (s) \\
    \hline
    2009 Dec 26 & 42.29 & 60 \\
    2009 Dec 30 & 46.25 & 60 \\
    2009 Dec 31 & 47.20 & 60 \\
    2010 Jan 01 & 48.16 & 60 \\
    2010 Jan 02 & 49.15 & 60 \\
    2010 Jan 03 & 50.15 & 60 \\
    2010 Jan 04 & 51.15 & 60 \\
    2010 Jan 07 & 54.20 & 60 \\
    2010 Jan 11 & 58.16 & 60 \\
    2010 Jan 12 & 59.14 & 60 \\
    2010 Jan 14 & 61.14 & 60 \\
    2010 Jan 16 & 63.13 & 60 \\
    2010 Jan 20 & 67.26 & 60 \\
    2010 Jan 25 & 72.20 & 60 \\
    2010 Jan 26 & 73.13 & 60 \\
    \hline
  \end{tabular}
  \label{tb:tb1}
\end{table}

\begin{figure}
  \centering
  \resizebox{\hsize}{!}{\includegraphics[angle=270]{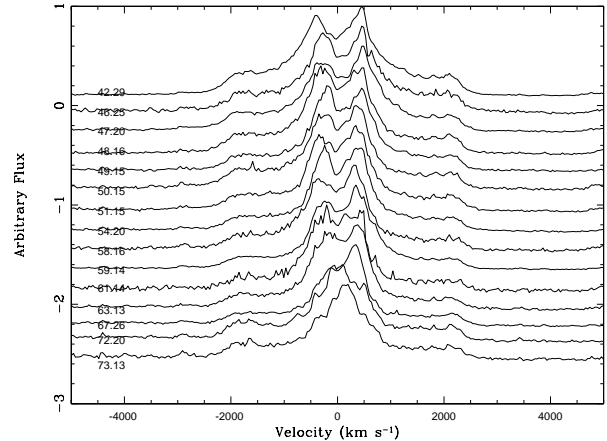}}
  \caption{H$\alpha$ emission line evolution. Numbers on the left are days post-outburst.}
  \label{fig1}
\end{figure}

Data reduction for FRODOSpec was performed through two pipelines. The first pipeline reduces the raw data. This includes bias subtraction, trimming of the overscan regions and flat fielding. Then the second pipeline performs a series of operations on the data, including flux extraction, arc fitting, throughput correction, linear wavelength rebinning and also sky-subtraction \citep{BSS12}. Any absorption features were patched over and then using the STARLINK Figaro IRFLUX task, correction for the instrumental efficiency/atmospheric transmission and relative flux calibration was performed on the KT~Eri spectra using the standard star HD19445. Figure~\ref{fig1} shows the evolution of the H$\alpha$ emission. The full spectra will be shown in a forthcoming paper.

\section{Modelling}\label{model}
Work by \citet{H72,S83,SOD95,GO00,HO03}, resolved nova shells with optical imaging which showed a myriad of structures. Several mechanisms for the formation of these structures were put forward. These included a common-envelope phase during outburst, the presence of a magnetised WD and an asymmetric thermonuclear runaway \citep[for a recent review see][]{OB08}. The most widely accepted shaping mechanism is that involving a common envelope phase.

Previous work, e.g. \citet[][hereafter MRB11]{MRB11} and \citet[][hereafter RDB11]{RDB11}, has extensively described the modelling procedures we use here and built a large grid of early outburst synthetic spectra. These synthetic spectra were derived from the following structures, i) polar blobs with an equatorial ring, ii) a dumbbell structure with an hour-glass over-density, iii) a prolate structure with an equatorial ring, iv) a prolate structure with tropical rings and v) a prolate structure with polar blobs and an equatorial ring. For each of the structures, the parameter space between 0 -- 90 degrees for the inclination angle (where an inclination $i$ = 90$^{\circ}$ corresponds to the orbital plane being edge-on, and $i$ = 0$^{\circ}$ being face-on) and 100 -- 8000 \kms\ for the maximum expansion velocity, were explored to retrieve a synthetic spectrum for each set of parameter combinations. The synthetic spectra were then compared with the observed spectra and flux matched via $\chi^2$ minimisation.

Here, we have also explored the optimizer module within {\sc shape}. This allows us to automatically improve the fit of a set of parameters using a least squares minimisation technique. Here we estimate an initial structure, velocity field (in order to retrieve a 1D line profile) and/or other parameters. The general strategy to optimise a structure is to choose one or more of the parameters to optimise. We then provide a range of values which the optimiser searches around. In general they are the same as those described above for inclination and maximum expansion velocity. These are taken as initial guesses by the module. Then, in the case of this study, a line profile is loaded which drives the minimisation process.

Once the optimisation process starts, the model is compared with the observation to generate a least-squares difference. The values of the parameters are varied over the range that was pre-set to find the least-squares solution and in the order they appear in the stack. The cycle of optimisation is repeated until a minimum difference is achieved.

This technique allows {\sc shape} to find the best fit results for the structure described above. The parameters that were allowed to be optimised were the inclination and maximum expansion velocity. Allowing for more parameters to vary is not advisable and might produce unrealistic models. However, this technique does not provide uncertainty estimates on the parameters at the time of writing this paper.

\begin{figure}
  \centering
  \resizebox{\hsize}{!}{\includegraphics[angle=270]{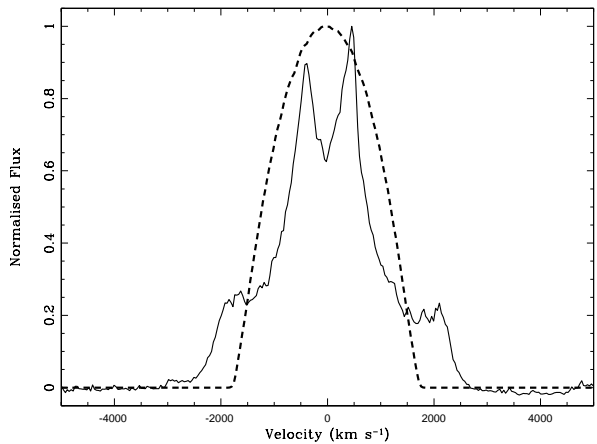}}
  \caption{Best fit synthetic spectra (dashed line), for a prolate structure with an equatorial ring, to the observed spectra (solid line) on day 42.29 following outburst.}
  \label{fig2}
\end{figure}

Applying the synthetic spectra derived from the structures described above to the observed spectra on day 42.29 following outburst did not provide a good fit to the observations (as an example, in Figure~\ref{fig2} we show the best fit result for a prolate structure with an equatorial ring). This is unsurprising due to the fact that the spectrum observed here is at a much later date than those in MRB11 and RDB11 (although in the latter case the structure was also evolved to a much later date). In particular RDB11, were able to demonstrate that when the early models (which were volume filled) were evolved to a much later date (in this case a thin shell), this appeared to replicate well the effect of the termination of the post-outburst wind phase and complete ejection of the envelope \citep{VPD02}.

\begin{figure}
  \centering
  \resizebox{\hsize}{!}{\includegraphics[trim=0.5cm 0cm 0cm 0cm, clip=true]{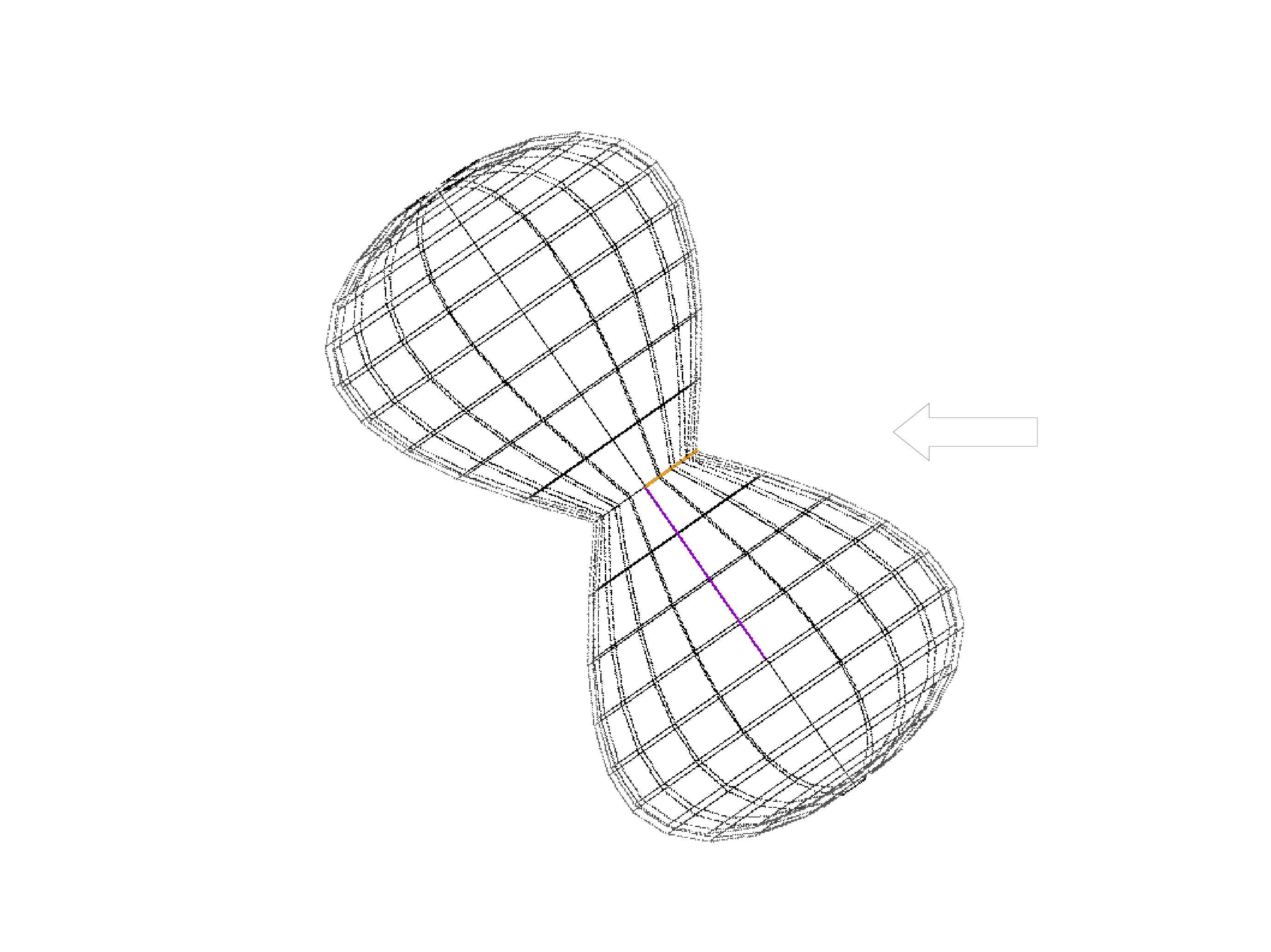}}
  \caption{KT Eri's model mesh, as the input geometry, as visualised in {\sc shape}. The inclination of the system is defined as the angle between the plane of the sky and the central binary system's orbital plane. The arrow indicates the observer's direction.}
  \label{fig3}
\end{figure}

As KT~Eri is suggested to harbour an evolved secondary \citep{JRD12}, we explored a structure similar to RS~Ophiuchi \citep[Figure~\ref{fig3}, see also][]{RBD09}. However, as mentioned in \citet{JRD12} the secondary in KT~Eri may not be as evolved as in RS~Oph. This suggests that there is not as strong a wind from the secondary pre-outburst, which may be associated with the origin of the central over-density modelled in RS~Oph.

The bipolar structure was assumed to have a ratio of the major to minor axis of 4:1. This was suggested by the ratio of the broad wing width to the central double peak separation being in this ratio (Figure~\ref{fig1}).

We further applied a radial density profile which varied as 1/$r$ (where $r$ is the radial distance away from the center of the outburst). This was investigated via multiple trial and error on the radial density profile, using the optimization module, to see what the best fit to the observed spectrum was, before exploring in more detail the parameter space of interest. The first spectrum explored was that of day 42.29 after outburst and then the best fit parameters found were then applied to day 63.13 after outburst, always keeping the parameters for the inclination and maximum expansion velocity the same.

As with previous work, bar \citet{RBD09}, to retrieve the synthetic line profiles we used the Mesh Renderer which directly used the 3D mesh structure (Figure~\ref{fig3}) and transfers it to a regular 3D grid. Along with the spatial distribution of the mesh, the physical properties that depend on position in space are transferred to the grid. These were the velocity, inclination and density. In the Mesh Renderer the emissivity, used to create the synthetic line profiles, is proportional to the density squared.

\section{Results}\label{results}
\subsection{Early epoch observations ($t$ = 42.29 days)}
\begin{figure}
  \centering
  \resizebox{\hsize}{!}{\includegraphics[width=130mm]{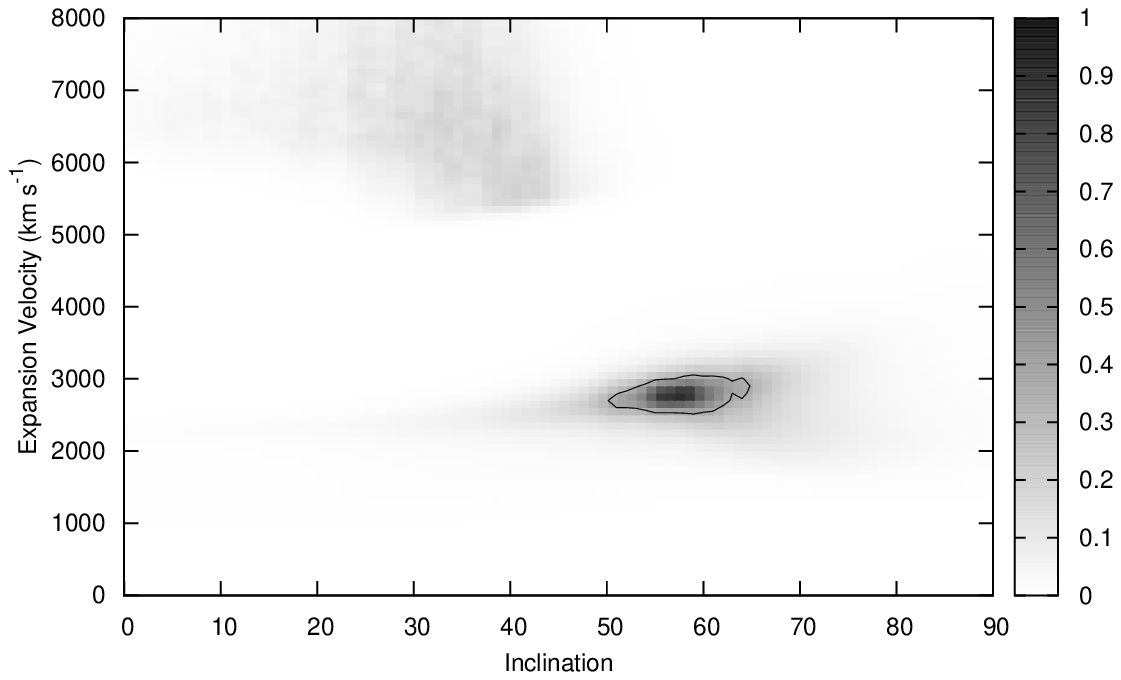}}
  \resizebox{\hsize}{!}{\includegraphics[angle=270, width=130mm]{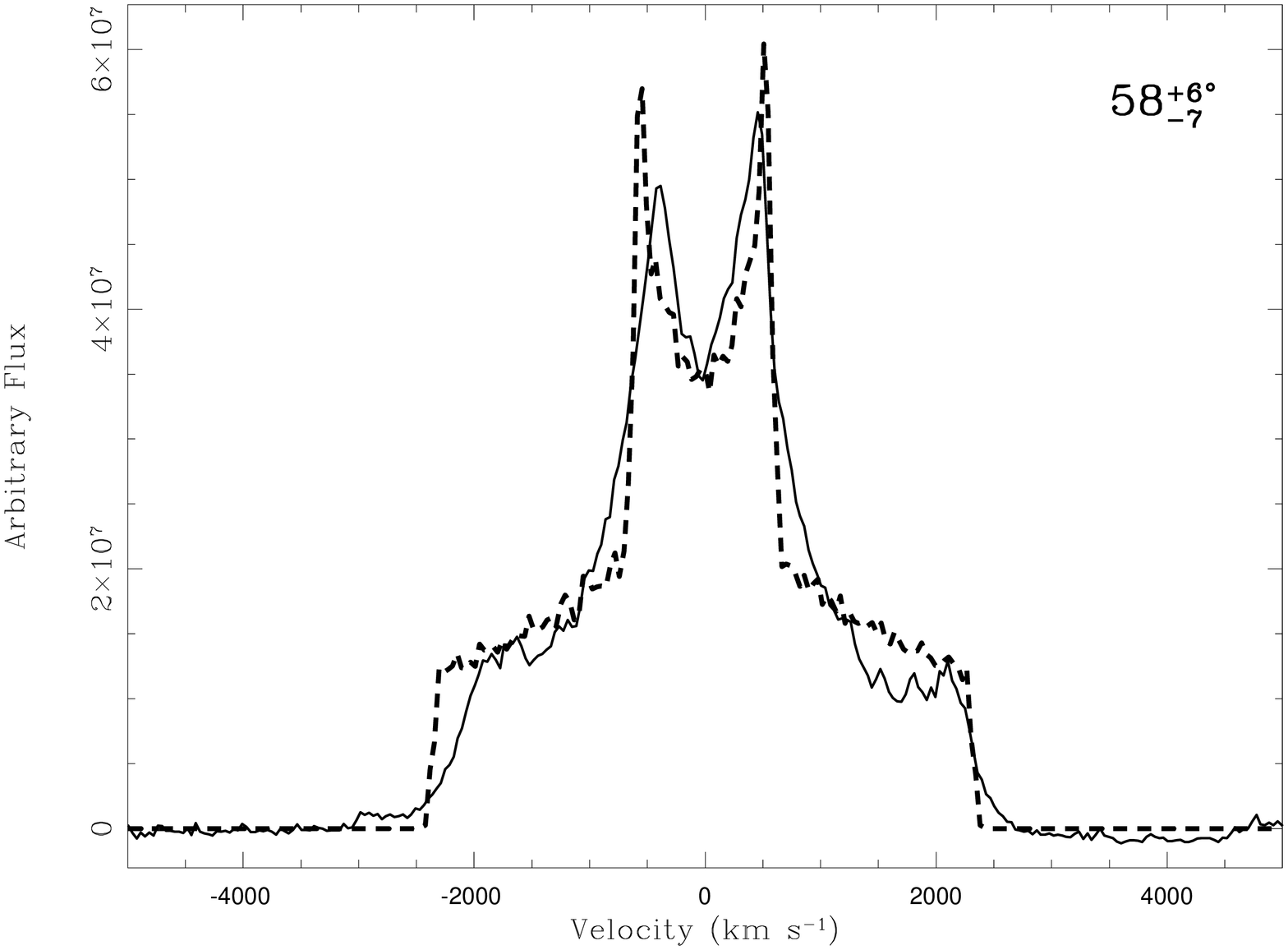}}
  \caption{Best-fit model results and comparison between the observed and model spectrum (day 42.29 after outburst) using the techniques described in MRB11 and RDB11. {\it Top} $-$ derivation of the most likely result for the structure, where the grey scale represents the probability that the observed $\mathcal{X}^2$ value is correct. The solid line represents the one sigma boundary. Furthermore, the island at the high velocity end is due to the finite resolution of the model \citep[for a brief discussion see ][]{R11}. {\it Bottom} $-$ The observed (solid black) and synthetic (dash black) spectra for the best-fit inclination, shown in the top right hand corner, with its respective one sigma errors.}
  \label{fig4}
\end{figure}

\begin{figure}
  \centering
  \resizebox{\hsize}{!}{\includegraphics[angle=270, width=186mm]{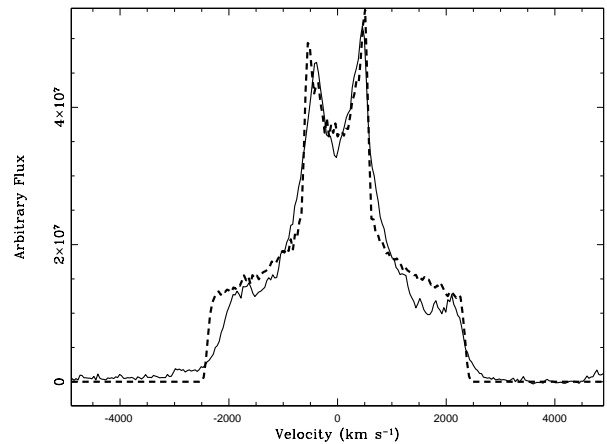}}
  \caption{Best-fit result using the Optimizer module technique assuming a 1/$r$ density distribution. The observed (solid black, $t$ = 42.29 days) and synthetic (dashed black) spectra are compared. The results suggest an inclination of 55\degree\ and $V_{\textrm{exp}}$ = 2800 \kms.}
  \label{fig5}
\end{figure}

The results of exploring the full parameter space are shown in Figure~\ref{fig4}. The best-fit result implies an inclination of $i$ = 58$^{+6}_{-7}$ degrees and a maximum expansion velocity $V_{\textrm{exp}}$ = 2800$\pm$200 \kms. In comparison, adopting the optimiser technique, the best-fit result suggests an inclination $i$ = 55\degree\ and $V_{\textrm{exp}}$ = 2800 \kms\ (Figure~\ref{fig5}), i.e. well within the uncertainty estimates of exploring the full parameter space.

\begin{figure}
  \centering
  \resizebox{\hsize}{!}{\includegraphics[angle=-90, width=130mm]{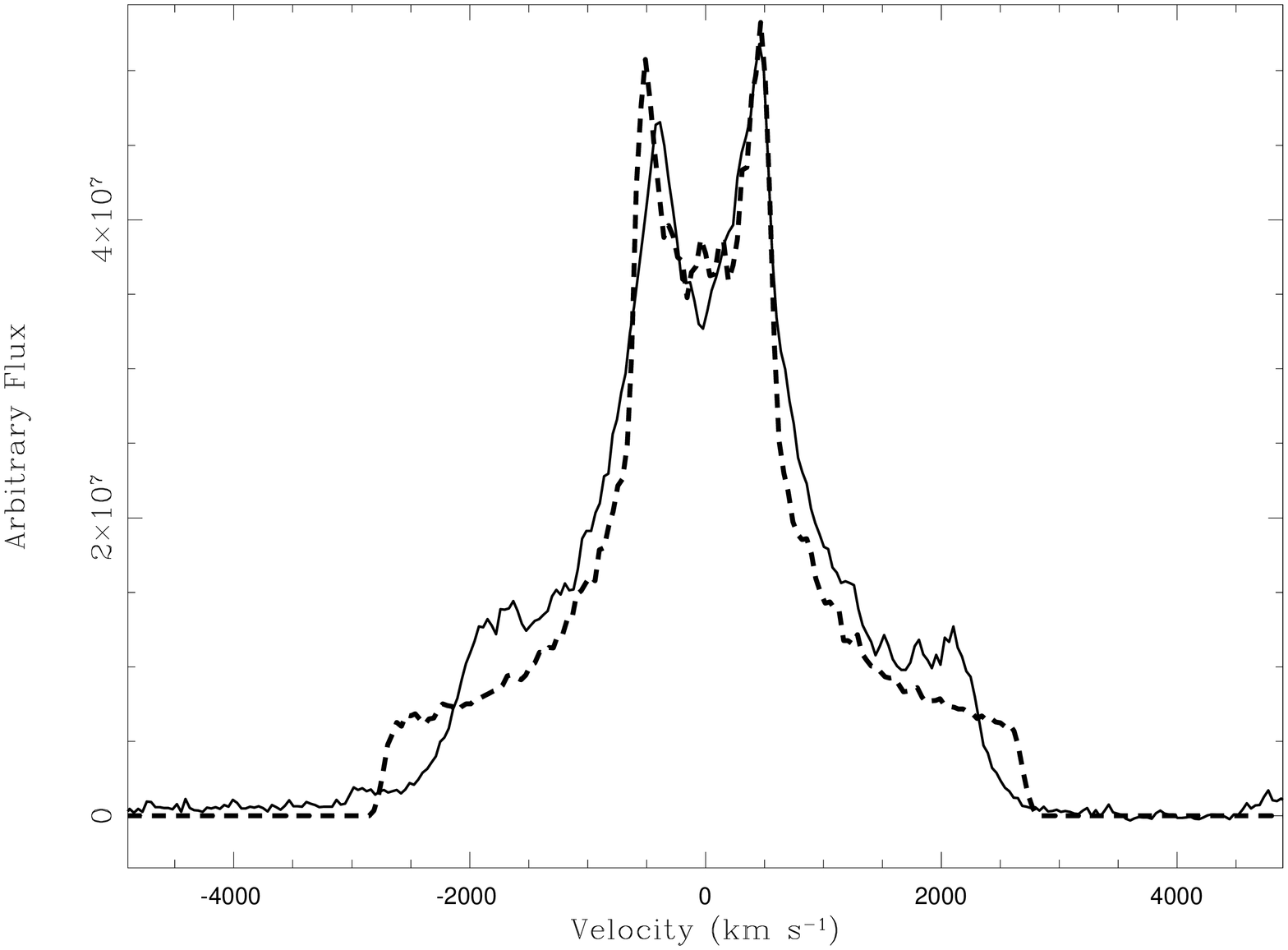}}
  \resizebox{\hsize}{!}{\includegraphics[angle=-90, width=130mm]{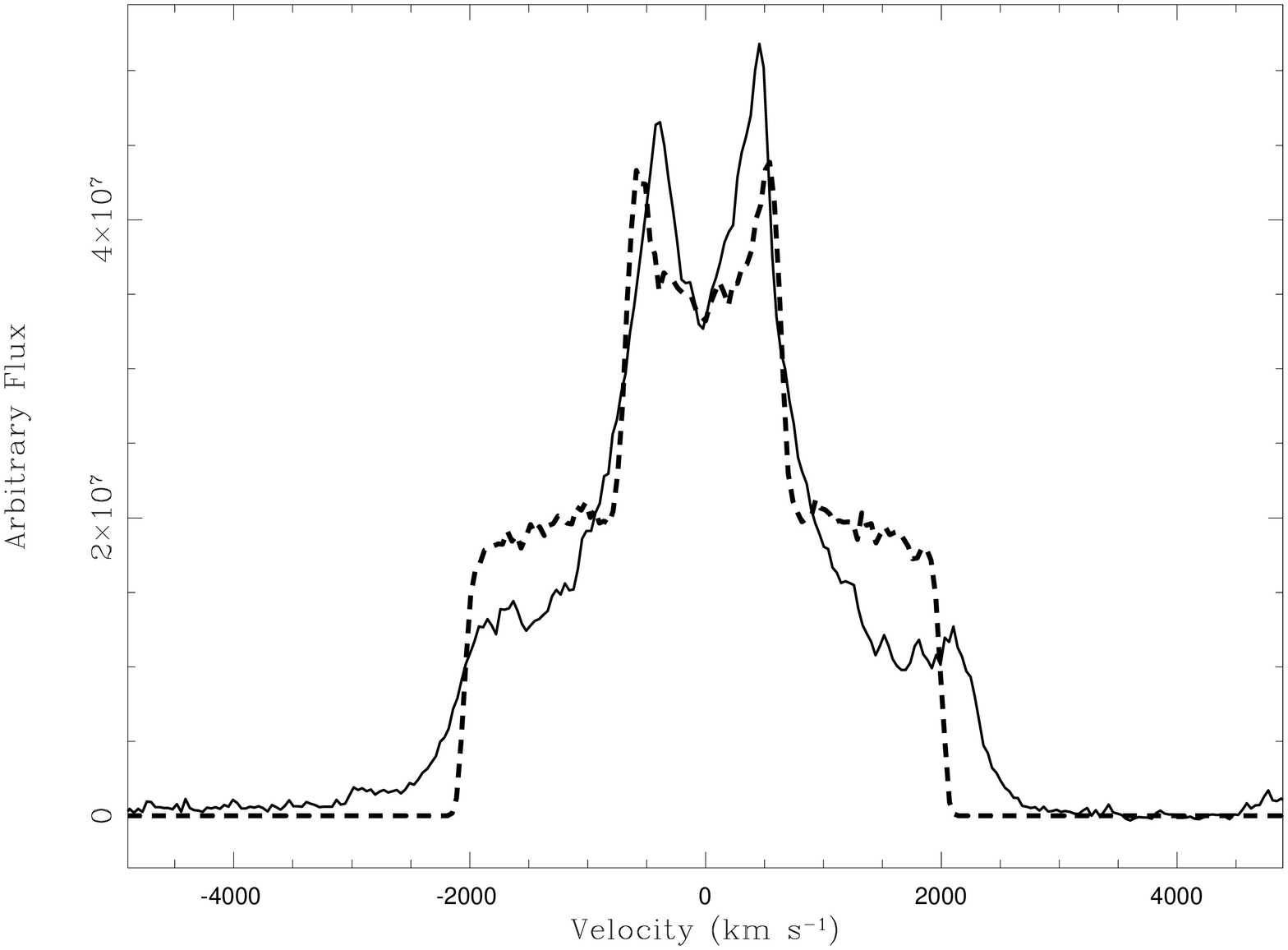}}
  \caption{Best-fit result using the Optimizer module technique for the earlier epoch observations ($t$ = 42.29 days). {\it Top} $-$ assuming a 1/$r^2$ radial density distribution, the results suggest an inclination of 51\degree\ and $V_{\textrm{exp}}$ = 3000 \kms. {\it Bottom} $-$ assuming a constant density distribution, the results suggest an inclination of 64\degree\ and $V_{\textrm{exp}}$ = 2600 \kms. The observed (solid black) and synthetic (dashed black) spectra are compared.}
  \label{fig6}
\end{figure}

The very similar results derived using the two methods allow us to very quickly explore different structures and find a best fit to the observed data using the optimizer technique. Considering the much reduced computational time that the optimizer technique allows for, we are able to explore other parameters efficiently. For example, in an attempt to see the effects of changing the radial density profile, we changed this parameter so that it varied as 1/$r^2$ and also assuming a constant density throughout (Figure~\ref{fig6}). The results suggest an inclination of 51\degree\ and 64\degree\ and maximum expansion velocities of 3000 \kms\ and 2600 \kms\ for a 1/$r^2$ and constant density profile, respectively. Figure~\ref{fig6} shows that these two assumptions for the density were not optimum fits to the observed spectra, either over or under predicting the fluxes in different parts of the spectrum. Therefore, the derived best-fit density profile is assumed to be that of a 1/$r$ distribution. However, the results for 1/$r^2$ and a constant density profiles are within the error bars derived for the 1/$r$ fit.

The degeneracy between our assumed structure and density profiles does not permit us to say with a degree of certainty which is the best fit density profiles; in all cases the structure reproduces the overall shape of the synthetic spectrum. Other effects, which are not taken in to account in the modelling, that may change the shape of the line profile include the consideration of self-absorption by the H$\alpha$ line, which in our models is assumed to be optically thin, but is consistent for example, with the fact that we observe structures in the line profiles that are to a large extent symmetrical.

\subsection{Later observations ($t$ = 63.13 days)}
Taking the values for the inclination and maximum expansion velocity from the previous subsection, both 1/$r$ and 1/$r^2$ density profiles were applied to the later epoch observations (Figure~\ref{fig7}). Inspection of Figure \ref{fig7} would suggest that the the profile results from a 1/$r^2$ density profile to be the better fit. However, the modelling of this line is complicated by the fact that there is an evolution of the line profile shape (Figure~\ref{fig1}). This evolution was discussed in \citet{R11}, who suggested that as the soft X-rays emerge, sampling nuclear burning on the surface of the WD, the H$\alpha$ emission arising from the ejecta may be contaminated by emission close to the central binary.

\section{Discussion and conclusions}\label{discussion}
The observed H$\alpha$ line profile from KT~Eri on day 42.29 modelled with a dumbbell shaped expanding nebulosity (Figure~\ref{fig3}) assuming a 1/$r$ radial density profile, $V_{\textrm{exp}}$ = 2800$\pm$200 \kms\ and inclination angle of 58$^{+6}_{-7}$ degrees as the best fit parameters. It is noteworthy that at this time the line profile we observe is likely to all be due to H$\alpha$ emission as the spectra in \citet{R11} do not show any contribution from [N~{\sc ii}] emission which played a role in the line shape of V2491 Cygni (RDB11). 
\begin{figure}
  \centering
  \resizebox{\hsize}{!}{\includegraphics[angle=-90, width=130mm]{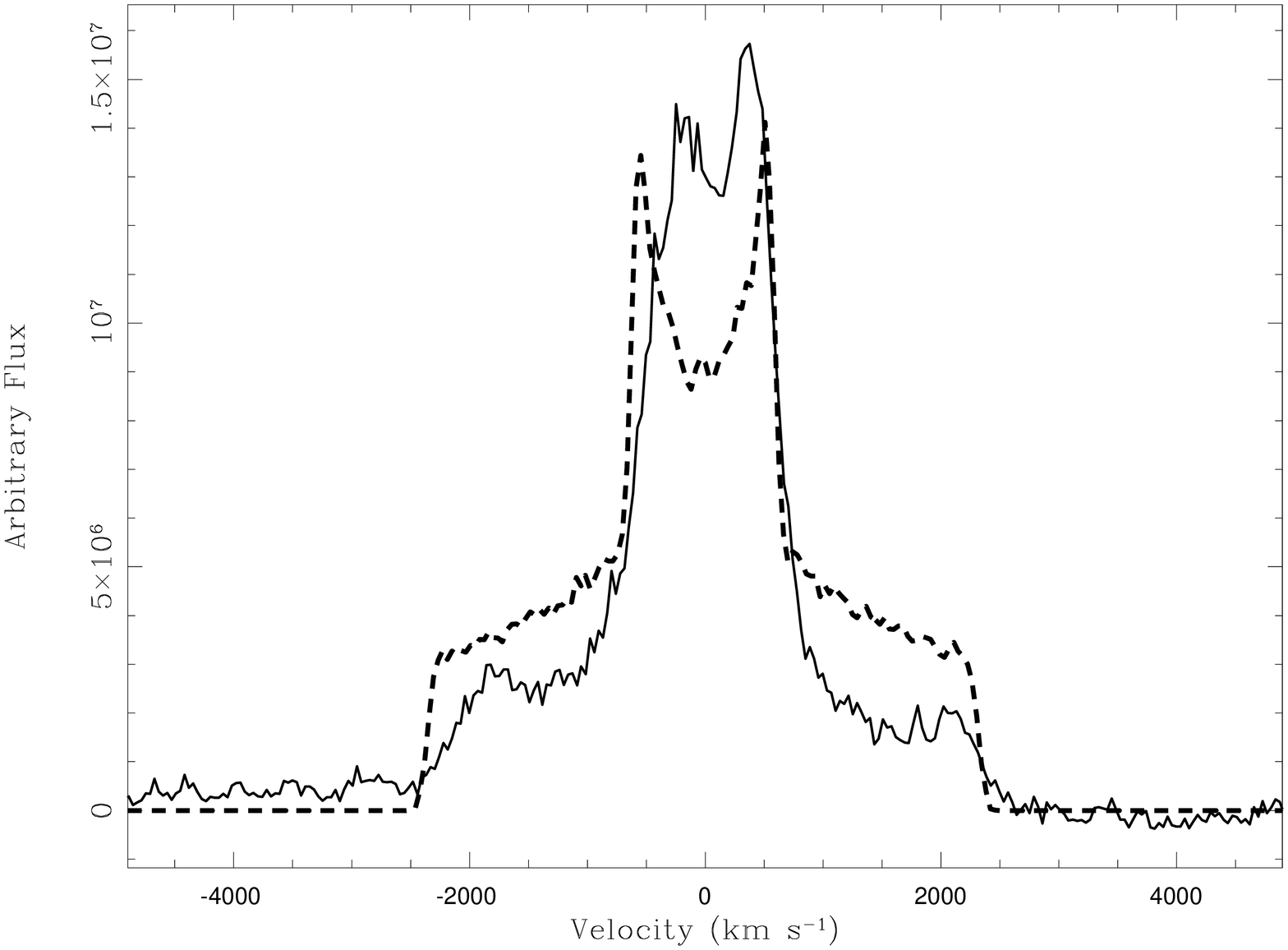}}
  \resizebox{\hsize}{!}{\includegraphics[angle=-90, width=130mm]{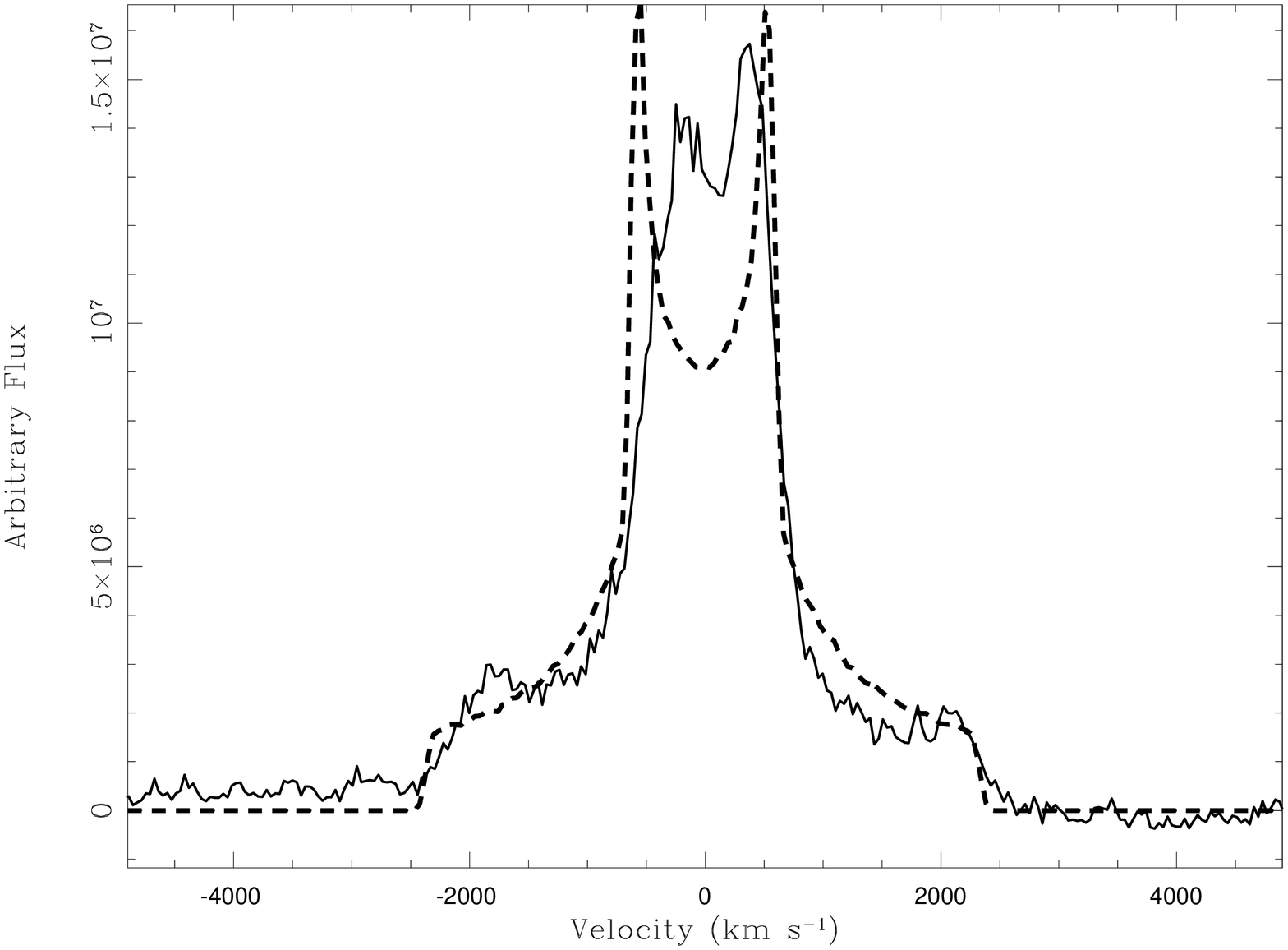}}
  \caption{Best-fit results for the later epoch observations ($t$ = 63.13) using the  Optimizer module technique assuming a 1/$r$ (top) and 1/$r^2$ (bottom) density profiles. The observed (solid black) and synthetic (dashed black) spectra are compared.}
  \label{fig7}
\end{figure}

The later epoch line profile (day 63.13)  was also modelled but this time the best fit resulted from a 1/$r^2$ density profile. However,  caution should be exercised here. \citet{R11} described the evolution of the H$\alpha$ emission line which showed the central regions of the line gain in intensity compared with the whole profile. This was interpreted as arising from material closer to the WD as the ejecta become optically thin and/or diffusion of the outer faster moving material. Similarly, evidence for emission by material closer to the WD may be supported by the relationship between the emergence of \heii\ 4686\AA\ and the Super-Soft Source \citep{R11}.

Earlier spectra than those shown here showed the line profile appears to be asymmetric with respect to the rest velocity \citep{R11}. This may be a combination of the systemic velocity of the system and/or the associated termination of the post-outburst wind phase, which happens during the first few weeks following the nova outburst \citep[e.g.][]{HKK00}, and complete ejection of the envelope \citep[][RDB11]{VPD02}, or optical depth effects.

What is certainly true with this type of study is that caution should be exercised when assuming a morphology arising from line profiles without the aid of imaging to constrain the models. Previous work dealt with modelling soon after the outburst, where most of the ejecta contribute to the line profile, and hence a constant density profile can be assumed; while at later stages although the morphology of the system will be the same, the main contributor to the line profile may be a combination of ejecta and UV illumination of the ejecta by the central source. This may be implied by the narrowing of the central peaks as shown in Figure~\ref{fig7} compared with Figure~\ref{fig6}. However, we cannot also rule out the fact that the higher velocity material will decrease in density faster as the ejected shell expands outward.

The structure and inclination angle derived here are consistent with those inferred by \citet{JRD12}. The inferred structure is also similar to that derived for RS Oph \citep{RBD09}. Here mass loss between outbursts from the red giant secondary star of the binary leads to an enhanced pre-outburst density in the binary equatorial plane external to the binary system itself. Hydrodynamic simulations by \citet{MP12} and \citet{MBP13} then produce the bipolar structures observed in the ejecta. In KT Eri, the secondary star is thought to be a low-luminosity giant and hence the mass of any pre-outburst envelope may be less than that in RS Oph. We also note however that there was no direct evidence of interaction with pre-existing material (e.g. through hard X-ray emission similar to that in RS Oph, \citealt{BOO06}, although the lower density of any pre-existing wind in KT Eri and the object's factor $\sim$4 greater distance than RS Oph would have made such emission less detectable in this case).

\section*{Acknowledgments}

The authors are grateful to W. Steffen and N. Koning for valuable discussion on the use of {\sc shape}. VARMR acknowledges financial support from the Royal Astronomical Society through various travel grants. The South African SKA Project is acknowledged for funding the postdoctoral fellowship position at the University of Cape Town. RMB was funded by an STFC studentship. We thank an anonymous referee for valuable comments on the original manuscript.

\label{lastpage}

\end{document}